\documentclass[twocolumn]{aastex631} %
\usepackage{CJK}
\usepackage{amsmath}
\usepackage{multirow}

\newcommand{\xJLS}{\exp[(\beta/\beta_\mathrm{crit})^{\alpha}-1]}
\newcommand{\flash}{{\tt FLASH}}
\newcommand{\mesa}{{\tt MESA}}
\newcommand{\Msun}{\mathrm{M_\odot}}
\newcommand{\Rsun}{\mathrm{R_\odot}}

\newcommand{\rT}{r_\mathrm{T}}
\newcommand{\tdyn}{t_\mathrm{dyn}}
\newcommand{\binit}{\beta_\mathrm{init}}
\newcommand{\rhocrhobar}{\rho_\mathrm{c}/\bar\rho}
\newcommand{\md}{\mathrm{d}}
\newcommand{\mbh}{M_\mathrm{BH}}

\graphicspath{{./}{figures/}}

\begin{document}
\begin{CJK*}{UTF8}{gbsn}
\title{Repeating Partial Tidal Encounters of Sun-like Stars Leading to their Complete Disruption}

\author[0000-0002-7866-4531]{Chang~Liu (刘畅)}
\affil{Department of Physics and Astronomy, Northwestern University, 2145 Sheridan Rd, Evanston, IL 60208, USA}
\affil{Center for Interdisciplinary Exploration and Research in Astrophysics (CIERA), Northwestern University, 1800 Sherman Ave, Evanston, IL 60201, USA}

\author[0000-0003-0381-1039]{Ricardo~Yarza}
\affil{Department of Astronomy and Astrophysics, University of California, Santa Cruz, CA 95064, USA}

\author[0000-0003-2558-3102]{Enrico~Ramirez-Ruiz}
\affil{Department of Astronomy and Astrophysics, University of California, Santa Cruz, CA 95064, USA}

\begin{abstract}
Stars grazing supermassive black holes on bound orbits may produce periodic flares over many passages, known as repeating partial tidal disruption events (TDEs). Here, we present 3D hydrodynamic simulations of sun-like stars over multiple tidal encounters. The star is significantly restructured and becomes less concentrated as a result of mass loss and tidal heating. The vulnerability to mass loss depends sensitively on the stellar density structure, and the strong correlation between the fractional mass loss $\Delta M/M_*$ and the ratio of the central and average density $\rho_{\mathrm{c}}/\bar\rho$, which was initially derived in disruption simulations of main-sequence stars, also applies for stars strongly reshaped by tides. Over multiple orbits, the star loses progressively more mass in each encounter and is doomed to a complete disruption. Throughout its lifetime, the star may produce numerous weak flares (depending on the initial impact parameter), followed by a couple of luminous flares whose brightness increases exponentially. Flux-limited surveys are heavily biased toward the brightest flares, which may appear similar to the flare produced by the same star undergoing a full disruption on its first tidal encounter. This places new challenges on constraining the intrinsic TDE rates, which need to take repeating TDEs into account. Other types of stars with different initial density structures (e.g., evolved stars with massive cores) follow distinct evolution tracks, which might explain the diversity of the long-term luminosity evolution seen in recently uncovered repeaters. 
\end{abstract}


\section{Introduction} \label{sec:intro}
The principal way in which supermassive black holes (SMBHs) have been studied observationally is via the emission from gas that slowly inspirals onto the SMBH. These steady-state active galactic nuclei (AGNs) are typically fed by gas that originates far from the SMBH itself, resulting in a feeding rate that varies little over decades \citep[e.g.,][]{2017ApJ...838..149A,2018ApJ...852...37A, 2019ApJ...883...31F, Dodd_2021,2021ARA&A..59...21G}. 
While they are magnificent laboratories for studying the behavior of matter in extreme environments, as the environmental conditions (e.g., accretion rates) of each system are fixed in time, we typically require demographic study of an ensemble of AGNs \citep[e.g.,][]{Merloni_2003,Falcke_2004,Wang_2006}. It can also be challenging to disentangle variations due to external quantities from variations due to SMBH properties. 
Furthermore, for the nearest SMBHs, including the one at the center of our own galaxy \citep[e.g.,][]{Narayan_1998}, the paucity of gas in the local Universe results in a tepid release of energy, hindering study of these dim SMBHs.

This behavior is in stark contrast to the frantic evolution experienced by an SMBH that has recently disrupted a star through tides \citep{Hills_1975,Rees_1988,Guillochon_2013}. When a star comes within a critical distance of an SMBH, immense tidal forces can remove a significant fraction, if not all, of the star's mass, resulting in a stream of debris that falls back onto the SMBH and powers a luminous flare lasting for months. The disruption of stars by SMBHs has been linked to more than a dozen optical/X-ray transients (tidal disruption events or TDEs) in the cores of galaxies out to $z\approx1$ \citep{2021ARA&A..59...21G,Velzen_2021,Andreoni_2022,Yao_2023}. The relatively short evolution timescale enables the study of the same SMBH subjected to different external conditions, which may transition between different modes of accretion \citep{Abramowicz_2013,Wevers_2021}. 

In the standard picture, the star is disrupted on a relatively weakly bound (nearly parabolic) orbit{, as is expected to be the dominant case in the classical scenario where stars are scattered to destructive orbits via two-body relaxation \citep{Stone_2020}}. Approximately half of the material removed from the star becomes bound to the SMBH \citep{Rees_1988}, which is distributed into many different orbits and falls back to the SMBH over a range of times  \citep{Evans_TDE_1989,Phinney_1989, 2012ApJ...760..103D, Guillochon_2013}. 
The rate that the material returns to the SMBH is characterized by three phases: a rapid rise to peak over a period of days, followed by relatively constant feeding over a period of weeks, and finally a power-law decay that can persist for decades -- these timescales mainly depend on the mass of the SMBH \citep{mockler_weighing_2019}. 

This picture can be dramatically altered if disrupted stars are placed on eccentric orbits (see \citealp{Rossi_2021} for a review). 
{The term ``eccentric TDE'' usually refers to TDEs on orbits with eccentricities below a critical value $e_\mathrm{crit}$, at which all the stellar debris remains gravitationally bound to the SMBH \citep{Hayasaki_2013}. In eccentric TDEs, the evolution of the accretion flow differs drastically from canonical parabolic TDEs. The debris falls back more rapidly, with a dramatic enhancement in the peak fallback rate \citep{Dai_TDE_2013, Hayasaki_2013, Hayasaki_2018}. The post-peak evolution of fallback rates is more complicated than a smooth $t^{-5/3}$ decline \citep{Park_2020, Cufari_2022a}. In contrast, TDEs on marginally eccentric orbits ($e_\mathrm{crit} < e < 1$) are more qualitatively similar to parabolic ones.}
Multiple dynamical processes can place stars in such eccentric orbits, including perturbations from massive objects (e.g., another SMBH) near the SMBH \citep{Chen_2011, 2023ApJ...959...18M, Melchor_2024}, the Hills breakup \citep{Hills_1988} of a tight stellar binary encountering an SMBH \citep{Cufari_2022, Lu_2023}, and tidal capture \citep{1992MNRAS.258..715K,1992MNRAS.255..276N,1995MNRAS.275..498D}.

Intriguingly, if the star is only partially disrupted (i.e., a substantial amount of the stellar mass retains self-bound) {on a bound orbit ($e<1$, but not necessarily below $e_\mathrm{crit}$), it can be repeatedly stripped as it} encounters the BH over and over again \citep{Kiroglu_2023, Liu2023}. {Stars on eccentric orbits around SMBHs may approach the tidal radius as they evolve toward the giant branch, when they would start feeding the SMBH every pericenter passage, and the episodic accretion events may contribute to the luminosity of quiescent SMBHs \citep{MacLeod_spoon_2013}. The repeating partial TDE scenario may also} explain some repeating nuclear transients recently uncovered in synoptic surveys, such as ASASSN-14ko \citep{Payne_2021, Payne_2022}, eRASSt\,J045650.3–203750 \citep{Liu_eRASSt_2023}, AT\,2018fyk \citep{Wevers_2019, Wevers_2023}, RX\,J133157.6–324319.7 \citep{Hampel_2022, Malyali_2023}, AT\,2020vdq \citep{Somalwar_2023}, Swift\,J023017.0+283603 \citep{Evans_2023, Guolo_2024}, and AT\,2022dbl \citep{Lin_2024}. All these candidates repeat/rebrighten at a timescale from months to years, and our ability to detect longer-period repeaters is strongly limited by the span of surveys. 

While most numerical work has focused on a single tidal encounter, the multiple-passage simulation space remains largely uncharacterized. Previous attempts to model a star surviving multiple tidal encounters rely on analytical recipes \citep{MacLeod_spoon_2013,2014ApJ...794....9M, Liu2023} or fitting formulas from hydrodynamic stellar libraries of single encounters \citep{Broggi_2024} for the star's response to tidal interaction. These approaches cannot characterize the star's tidal deformation and the mass loss coherently over time. {A fully hydrodynamic approach was first adopted in \citet{Guillochon_planet_2011}, to model repeating tidal encounters of a giant planet orbiting a star, yet the mass ratio ($\sim$$10^3$) of the two objects is much less extreme than typical TDEs ($\sim$$10^6$).} In this {paper}, we present 3D hydrodynamic simulations of multi-orbit encounters between a sun-like star and an SMBH, in which we track the structural evolution of the star and the long-term trend of mass loss over multiple encounters. 

The {paper} is organized as follows. In Section~\ref{sec:simulation} we elaborate the necessity of hydrodynamic simulations and our model setups.
In Section~\ref{sec:result} we present the evolution of the star's density structure over time and the amount of mass removed in each tidal encounter. This is followed by our discussion of the implications of the long-term trend of luminosity in these repeaters in Section~\ref{sec:flare}. Finally, in Section~\ref{sec:discussion} we discuss how the initial structure of the star affects the behavior of the observed repeater and address the challenges of inferring the TDE rates both theoretically and observationally, with the existence of an underlying population of repeating partial TDEs. {These discussions are followed by a summary of the major assumptions and simplifications in modeling repeating TDEs and the potential systematics in our interpretation.} We draw our conclusions in Section~\ref{sec:conclusion}.

\section{Simulating Multiple Tidal Encounters}\label{sec:simulation}
\subsection{Hydrodynamic Simulations}

In both full and partial TDEs, the mass fallback rate $\dot M$ onto the SMBH depends on the distribution of the orbital energy, $\md M/\md E$, within the debris tail bound to the SMBH \citep{Rees_1988, Evans_TDE_1989, Phinney_1989,Guillochon_2013},
\begin{equation}\label{eq:dotM}
    \dot M=\frac23\frac{\md M}{\md E}\left(\frac{\pi^2G^2\mbh^2}{2}\right)^{1/3}t^{-5/3}.
\end{equation}
The energy distribution in the debris and thus $\dot M$ depend sensitively on the mass and age of the star, as well on its predisruption orbital properties \citep[e.g,][]{Law-Smith2019, Law-Smith2020}. Hydrodynamic simulations are required to handle both the fluid dynamics within the debris tail and the gravitational interaction with the surviving core \citep{2008ApJ...679.1385R,2009ApJ...695..404R,Lodato_2009,Ramirez-Ruiz_2009, Coughlin_2019, Miles_2020}. The structure of the surviving star is also strongly deformed, which governs its behavior in subsequent encounters. Hydrodynamic simulations have to be employed to model the star's response to mass loss \citep{MacLeod_spoon_2013, Ryu_2020c}, tidal excitation that deposits energy to the stellar interior \citep{Li_2013, Manukian_2013}, and the re-accretion of marginally bound material as the star recedes from the pericenter to a region with a weaker tidal field \citep{2005Icar..175..248F,2011ApJ...731..128A,Guillochon_planet_2011}. In this work, our simulations focus on the evolution of the star's structure following a series of tidal encounters.

We follow the approach of \cite{Liu2023} and \citet{Law-Smith2019,Law-Smith2020} to simulate a sun-like star over multiple orbits. The initial density profiles of a sun-like star are modeled with the Modules for Experiments in Stellar Astrophysics \citep[\mesa, version r23.05.1;][]{Paxton2011, Paxton2013, Paxton2015, Paxton2018, Paxton2019, Jermyn2023}, using the same setup presented in \citet{Law-Smith2019}. As an outcome, the stellar radius $R$ is 1.03\,$\Rsun$ at an age of 44.5\,Gyr. The profile is then mapped to a 3D adaptive mesh refinement hydrodynamics code \flash\ \citep{Fryxell_FLASH_2000}. The adiabatic index of the gas is $\Gamma=5/3$. We adopt a $10^6\,\Msun$ SMBH but the results are scalable to other $\mbh$. The major difference from the setups of \citet{Liu2023} and \citet{Law-Smith2019, Law-Smith2020} is the much smaller domain size adopted (100\,$\Rsun$ versus 1000\,$\Rsun$). Our zoom-in simulations prioritize the resolution within the star at the expense of not modeling the extended tidal tails. 

We study orbits with three initial impact parameters, $\binit=0.5$, 0.6, and 1.0, where $\beta$, the ratio of the tidal radius $\rT\equiv R_*(\mbh/M_*)^{1/3}$ and the pericenter distance $r_\mathrm{p}$, characterizes how deep the star penetrates the tidal radius. These values are well below the critical $\beta_\mathrm{crit}$ for a full disruption \citep{Mainetti_2017,Law-Smith2020, Ryu_2020b}, so the star can survive at least a couple of orbits. We also ensure that all the encounters are nonrelativistic ($r_\mathrm{p}\gtrsim50\,r_\mathrm{g}$, whereas we expect non-negligible relativistic effects when $r_\mathrm{p}\lesssim 10\,r_\mathrm{g}$; e.g., \citealp{1993ApJ...410L..83L,Cheng_2014,Servin_2017,Tejeda_2017,Gafton_2019,Stone_2019,Ryu_2020d}). We begin the simulation by relaxing the star onto the grid for $5\,\tdyn$ (the dynamical timescale $t_\mathrm{dyn}=1664$\,s), $10\,\rT$ away from the SMBH before starting the eccentric orbital evolution. For the $\binit=0.6$ and $1.0$ models, the eccentricity $e$ is set to be 0.9, so the orbital periods $P_\mathrm{orb}$ are $\approx$400\,$\tdyn$ and $\approx$200\,$\tdyn$ each, corresponding to a couple of days. For the $\beta=0.5$ model, we adopt a smaller $e=0.8$ for efficiency, and the period is same as that of the $\beta=1.0$ model. {We note that in reality, $P_\mathrm{orb}$ is usually much longer ($\gtrsim$100\,days in most of the repeaters discovered, corresponding to $e\gtrsim0.99$).} 
\citet{Liu2023} have shown that $\md M/\md E$ is not significantly different for disruption on highly eccentric ($e\gtrsim 0.9$) or parabolic orbits ($e=1$), {so by simulating the star on low-$e$ orbits we can approximate the star's tidal response on higher-$e$ orbits with the same $\beta$ at a low computational cost.} {The caveat is that} the star in the $\binit=0.5$ model {($e=0.8$) probably experiences} stronger tidal effects than the same star on an orbit with the same $\binit$ but a longer $P_\mathrm{orb}$.

We stop the simulations when: (i) $\Delta M/M_*>50\%$, when {the star is essentially completely disrupted and} we usually expect the orbit of the remnant to be strongly altered \citep{Gafton_2015}; or (ii) {the total number of blocks exceeds 1.2 million.\footnote{At this stage, simulating an additional passage becomes unaffordable ($>$20,000 CPU hours) without lowering the resolution.}} We calculate the specific self binding energy $E_*$ in an iterative approach following Equations~(2) and (3) of \citet{Guillochon_2013}. 
The remaining stellar mass $M_*$ is the sum of mass in cells with $E_{*}<0$. The mass loss $\Delta M$ in each passage is defined as the decrease in $M_*$ in one orbit (as the star returns to $r=10\,\rT$), unless $M_*\to0$ beforehand. In practice, we find $\Delta M$ stabilizes within $\approx$100\,$\tdyn$ after the previous encounter, and the variation in $\Delta M$ afterward is $\lesssim$$10^{-6}\,\Msun$ per $\tdyn$ before the next encounter.

\begin{figure*}
    \centering
    \includegraphics[width=\linewidth]{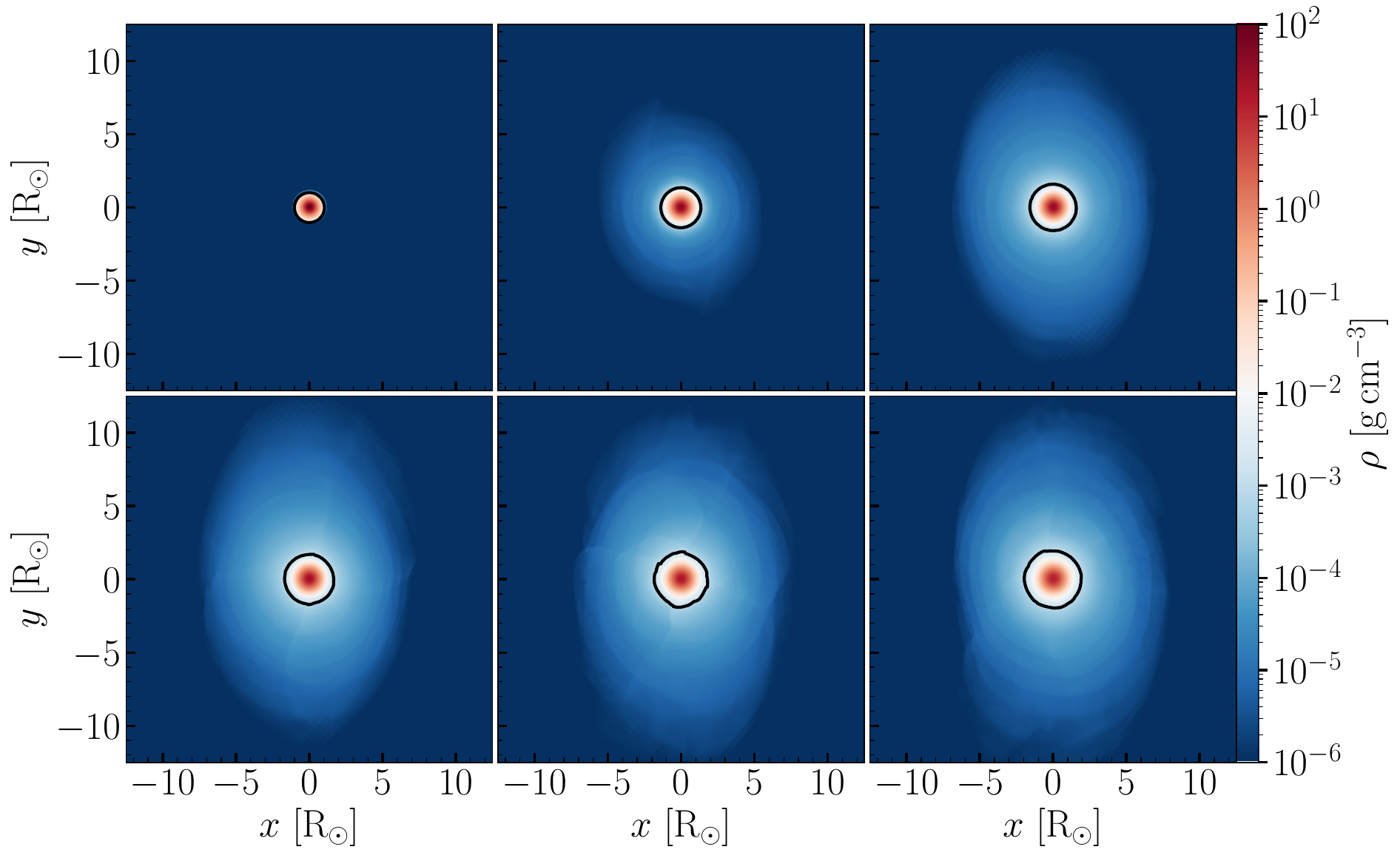}
    \caption{Density maps of the star on the $\beta_\mathrm{init}=0.6$ orbit before each of its six pericenter passages ($10\,\rT$ from the SMBH). The black contour denotes the density threshold $\rho=10^{-3}$\,g\,cm$^{-3}$, which we define as the stellar surface. The star expands over time and $\rho_\mathrm{c}$ drops. The stellar surface is continuously heated by the re-accretion of the marginally bound materials, which has a much greater $\rT$ and will be easily removed in the subsequent encounter.}
    \label{fig:snapshots}
\end{figure*}

At the end of our simulations, {the $\binit=1.0$ model retains $<$$50\%$ of its original mass after three passages, while in the $\binit=0.5$ and $0.6$ models, most of the material remains bound to self-gravity.}
The stellar parameters at each passage are listed in Table~\ref{tab:multi_pass}. In Figure~\ref{fig:snapshots} we show snapshots of the star in the $\binit=0.6$ model, when the orbital separation is $10\,\rT$ before each of its six pericenter passages. The central density $\rho_\mathrm{c}$ of the star decreases over time as a result of tidal energy injection, and the star expands. A diffuse envelope of marginally bound material is developed after the first encounter, which also keeps heating the stellar surface via re-accretion \citep{Guillochon_planet_2011, MacLeod_spoon_2013}. {But the total mass of this envelope, presented as the uncertainty of the $\Delta M$ in Table~\ref{tab:multi_pass}, is tiny ($\lesssim$0.1\% of the remaining stellar mass). In addition, since it extends to a couple of stellar radii, corresponding to a much greater effective $\beta$, the envelope formed in one tidal encounter is easily removed in the subsequent one, leaving essentially no cumulative effects. Up to $\sim$10\% of the peeled mass $\Delta M$ comes from this marginally bound envelope, which will slightly alter the mass fallback rate (and probably the consequent luminosity), but only at very early times before the fallback of the bulk stripped mass.} 

\begin{deluxetable*}{c|c|ccccc}
\tablecaption{Stellar Parameters in Each Encounter and the Corresponding Mass Loss.\label{tab:multi_pass}}
\tablehead{
\multirow{2}{*}{$\binit$} & \multirow{2}{*}{$N_\mathrm{passage}$} & $R_*$ & $\rho_\mathrm{c}$ & $\rhocrhobar$ & $\xJLS$ & $\Delta M$ \\& & $(\Rsun)$ & ($\mathrm{g\,cm^{-3}}$) & & & $(10^{-2}\,\Msun)$
             }
\startdata
    \multirow{10}{*}{0.5}& 1$^\dagger$ & $1.03\pm0.01$ & 143.7 & $109\pm3$ & $0.758\pm0.001$ & $0.001\pm0.015$ \\
                         & 2$^\dagger$ & $1.21\pm0.01$ &  56.9 & $71\pm2$  & $0.774\pm0.001$ & $0.14\pm0.05$ \\ 
                         & 3$^\dagger$ & $1.33\pm0.03$ &  42.7 & $70\pm4$  & $0.787\pm0.003$ & $0.37\pm0.09$\\ 
                         & 4           & $1.40\pm0.05$ &  35.8 & $70\pm8$  & $0.795\pm0.006$ & $0.57\pm0.13$\\ 
                         & 5           & $1.43\pm0.05$ &  31.2 & $65\pm6$  & $0.798\pm0.005$ & $0.77\pm0.17$\\ 
                         & 6           & $1.45\pm0.05$ &  27.8 & $61\pm6$  & $0.801\pm0.006$ & $0.99\pm0.20$\\ 
                         & 7           & $1.47\pm0.05$ &  25.0 & $57\pm6$  & $0.804\pm0.005$ & $1.31\pm0.23$\\ 
                         & 8           & $1.50\pm0.06$ &  22.6 & $56\pm6$  & $0.808\pm0.006$ & $1.66\pm0.26$\\ 
                         & 9           & $1.53\pm0.07$ &  20.6 & $55\pm8$  & $0.812\pm0.007$ & $2.09\pm0.29$\\ 
                         & 10          & $1.55\pm0.07$ &  18.7 & $53\pm8$  & $0.817\pm0.007$ & $2.71\pm0.29$\\
     \hline
     \multirow{6}{*}{0.6}& 1$^\dagger$ & $1.03\pm0.01$ & 143.7 & $109\pm3$ & $0.779\pm0.001$ & $0.19\pm0.07$\\
                         & 2           & $1.34\pm0.02$ &  41.9 & $71\pm4$  & $0.815\pm0.002$ & $1.32\pm0.35$\\ 
                         & 3           & $1.51\pm0.06$ &  29.9 & $74\pm9$  & $0.836\pm0.005$ & $3.05\pm0.62$\\ 
                         & 4           & $1.59\pm0.09$ &  23.1 & $69\pm11$ & $0.849\pm0.007$ & $5.41\pm0.64$\\ 
                         & 5           & $1.71\pm0.10$ &  17.8 & $70\pm12$ &
                         $0.866\pm0.007$ & $9.46\pm0.73$\\
                         & 6           & $1.84\pm0.11$ &  13.1 & $72\pm13$ &
                         $0.887\pm0.006$ & $15.40\pm0.65$\\
     \hline
     \multirow{3}{*}{1.0}& 1           & $1.03\pm0.01$ & 143.7 & $109\pm3$ & $0.846\pm0.001$ & $8.10\pm0.08$\\
                         & 2           & $1.72\pm0.13$   &  50.4 & $199\pm44$& $0.920\pm0.005$ & $21.55\pm0.19$\\ 
                         & 3       & $1.97\pm0.15$   &  19.9 & $153\pm35$& $0.965\pm0.003$ & $56.74\pm0.30$\\
\enddata
\tablecomments{Passages where there is unresolved numerical tidal dissipation are labeled by $\dagger$.}
\end{deluxetable*}

\subsection{Adiabatic Mass Loss Approximation}

Characterizing a star's response to mass loss is critical in understanding the stability of mass transfer in binary systems. When the mass loss occurs at a timescale much shorter than the thermal timescale at the stellar surface, a common assumption is that the the star would react adiabatically within $\tdyn$ \citep[e.g.,][]{Hjellming_1987, Soberman_1997}. As a generalized interacting binary system, the long-term evolution of a star undergoing repeating partial TDEs has also been explored assuming adiabatic mass loss \citep{MacLeod_spoon_2013, 2014ApJ...794....9M, Liu2023}. In this work, we also set up an adiabatic mass loss model in comparison with the fully hydrodynamic treatment, in order to quantify the importance of tides.

We start with the same \mesa\ sun-like star model\footnote{{Our \mesa\ inlists and output files are available on Zenodo at \dataset[doi:10.5281/zenodo.14084158]{https://doi.org/10.5281/zenodo.14084158}}.} and remove a total mass of $0.1\,\Msun$ from its envelope via stellar wind at a constant rate of $10^{-6}\,\Msun\,\mathrm{yr^{-1}}$. The corresponding mass loss timescale is $10^5\,$yr, significantly lower than the thermal (Kelvin-Helmholtz) timescale ($\approx$$10^7$\,yr) but much longer than the dynamical timescale ($\approx$hr). Throughout the evolution, the entropy profile remains nearly unchanged, except for that in the superadiabatic stellar surface \citep[see also][]{Woods_2011}. While the mass loss rate seems to be specified here, the result is not sensitive to it -- bumping up the mass loss rate ($10^{-5}\,\Msun\,\mathrm{yr^{-1}}$) has a minimal impact on the stellar structure when the same amount of total mass is removed.

\subsection{Tidal Dissipation}\label{sec:tidal_dis}
Tidal forces alter the structure of the star by converting orbital energy into mechanical energy stored in oscillations. The energy deposited into these oscillations is roughly proportional to the square of their amplitude \citep{Guillochon_planet_2011}, $|{\delta E_*}/{E_*}| \approx (\delta R/R)^2$, where $|E_*|\simeq GM^2/R$ is the magnitude of the gravitational binding energy of the star. For weak tidal encounters in which $\delta R$ is below the resolution of the simulations, tides will be dissipated numerically, puffing up the star in an artificial way. In our simulations, the typical cell size within the star is $\Delta L=0.025\,\Rsun$. The energy deposition during a single passage is \citep{Press_1977}
\begin{equation}\label{eq:deltaE}
    \delta E_* = \left(\frac{GM^2}{R}\right)\beta^6T(\beta),
\end{equation}
where $T(\beta)$ also depends on the stellar structure. For a $\gamma=4/3$ polytrope, a good approximation of sun-like stars, $T(\beta)\approx10^{-2}$--$10^{-1}$ for $\beta\approx0.5$--1.0 \citep{Ivanov_2001}. For $\binit=0.5$, in the first passage, we have $|\delta E_*/E_*|\approx 10^{-4}$, requiring a cell size finer than $\approx$$0.01\,\Rsun$, beyond our computational ability in multi-orbit simulations. This means we inevitably overestimate the energy deposition at the beginning of the simulation.

As star expands during subsequent passages, both $R_*$ and $\beta$ increase, easing the resolution requirement to resolve the tides. When a sun-like star has expanded to $R_*$ over multiple passages, our cell size $\Delta L$ must satisfy
\begin{align*}
    \frac{\Delta L}{\Rsun} & \lesssim \beta^{3}\left(\frac{R_*}{\Rsun}\right)T^{1/2}(\beta) \\
    & \simeq \binit^3\left(\frac{R_*}{\Rsun}\right)^4T\left[\binit\left(\frac{R_*}{\Rsun}\right)\right].
\end{align*}
By adopting the analytical fit to $T(\beta)$ in \citet{Generozov_2018}, we solve $R_{\Delta L}$, the minimum $R_*$ to which the star must expand such that $\Delta L\simeq\delta R$, to be $\approx$1.4\,$\Rsun$, 1.2\,$\Rsun$, and 0.75\,$\Rsun$ for $\binit=0.5$, 0.6, and 1.0. In Table~\ref{tab:multi_pass} we mark the corresponding orbits with insufficient resolution.
We will discuss this further in Section~\ref{sec:result}.

We note that the numerical dissipation of the hydrodynamic scheme is the only dissipation mechanism in our simulations. Once $R_*$ exceeds $R_{\Delta L}$, we do not observe significant decay of the $l=2$ modes between pericenter passages. Consequently, the tidal modes remain excited and interfere with newly excited modes in the subsequent encounter \citep{Mardling_1995a, Mardling_1995b}. The response of the star during the following passage depends on the phase of the oscillations at pericenter, leading to chaotic behavior. Therefore it is impossible to predict the exact number of orbits a star could survive before the complete disruption, which can be sensitive to small perturbations of orbital parameters \citep[e.g., $\binit$ and eccentricity;][]{Guillochon_planet_2011}. In realistic stars, the oscillation energy can be transferred from the primary $l=2$ mode to higher-order daughter modes via nonlinear coupling \citep{Kumar_1996, Weinberg_2012}, which can then be efficiently damped (e.g., via turbulence) within the realistic orbital timescale ($\gtrsim$yr) of repeating TDEs, reducing the degree of chaos. Nevertheless, tidal dissipation remains an open problem and is a source of uncertainty in our calculations.

\section{The Star's Response to Repeating Partial Disruptions}\label{sec:result}

\begin{figure*}
    \centering
    \includegraphics[width=\textwidth]{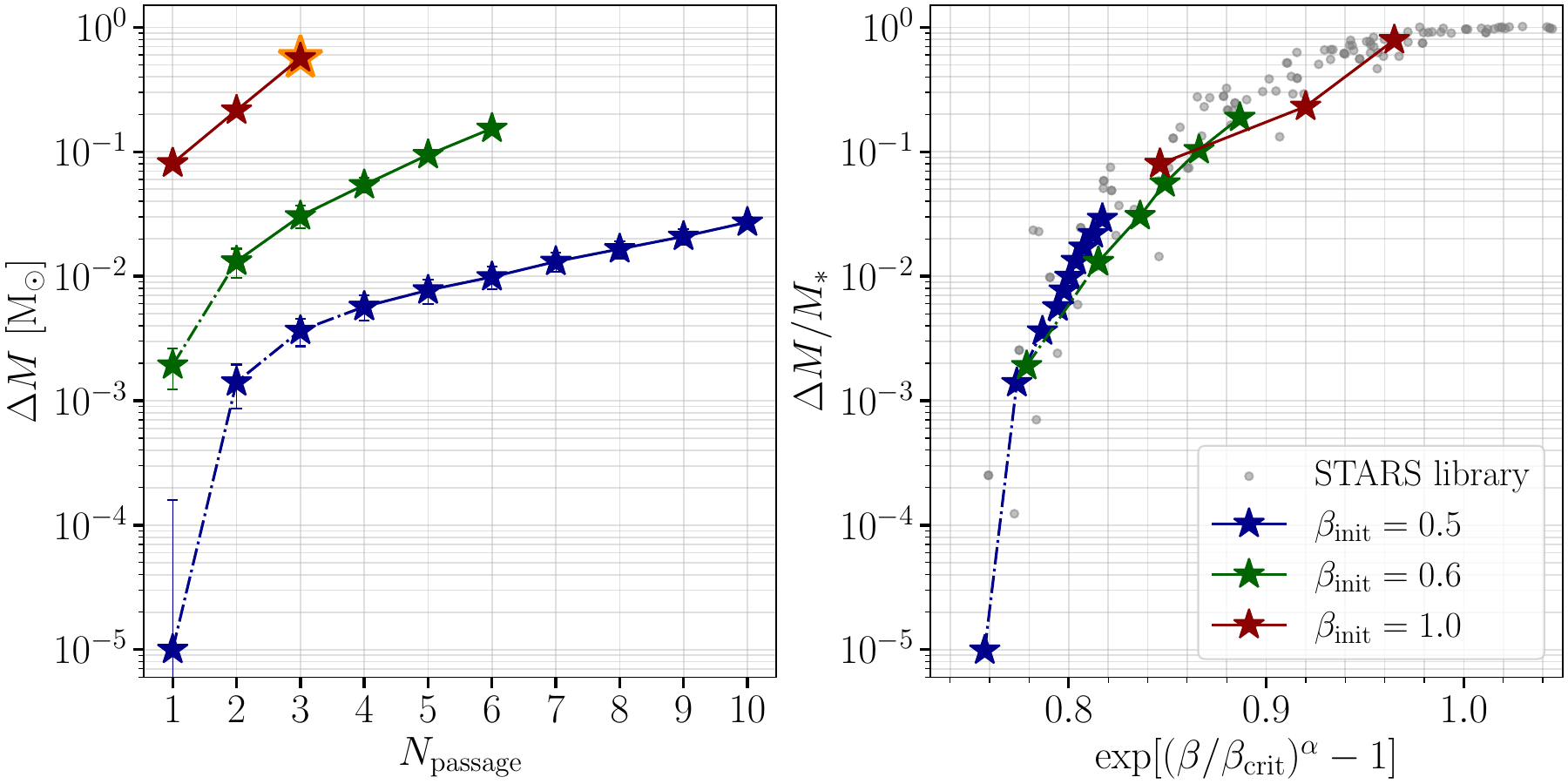}
    \caption{The mass $\Delta M$ unbound by the SMBH for three different $\beta_\mathrm{init}$. {\it Left:} $\Delta M$ at each passage. The uncertainty is given by the mass that is marginally bound to the star ($E_*<0$, $\rho < 10^{-3}\mathrm{g\,cm^{-3}}$). {The orange marker indicates the star is almost completely disrupted, having lost $>$50\% of its mass.} The dashed-dotted lines indicate that numerical tidal dissipation dominates stellar evolution in the corresponding orbits. {\it Right:} fractional mass loss $\Delta M/M_*$ as a function of $\xJLS$ (see the text for the definition). Overplotted are the results from the STARS library in \citet{Law-Smith2020}, suggesting $\xJLS$ well quantifies the vulnerability of the star to mass loss.}
    \label{fig:deltaM}
\end{figure*}
In partial TDEs, the fractional mass loss $\Delta M/M_*$ depends on the stellar structure. \citet{Ryu_2020c} showed that $\Delta M/M_*$ can be quantified with $\beta$ and a physical tidal radius, $\mathcal{R}_\mathrm{t}$. Unlike $\rT$, which depends on the average density of the star $\bar\rho$, $\mathcal{R}_\mathrm{t}$ depends on the ratio of the central density and the average, $\rhocrhobar$ \citep{Ryu_2020a, Ryu_2020b}. Similarly, \citet{Law-Smith2020} defined
\begin{equation}\label{eq:x_JLS}
x\equiv\xJLS,
\end{equation}
which also strongly correlates with $\Delta M/M_*$. Here, $\alpha\equiv\left(\rhocrhobar\right)^{-1/3}$ and $\beta_\mathrm{crit}$ is the critical impact parameter for full disruption, again depending on $\rhocrhobar$:
\begin{equation}
    \beta_\mathrm{crit}\simeq
    \left\{
    \begin{array}{cc}
        0.5(\rhocrhobar)^{1/3}, & \rhocrhobar\lesssim 500,  \\
        0.39(\rhocrhobar)^{1/2.3}, & \rhocrhobar\gtrsim500. 
    \end{array}
    \right.
\end{equation}
Here, we show that the relation between $\xJLS$ and $\Delta M/M_*$ still holds even if the star has a profound tidal interaction history, with its density structure being substantially perturbed.

Definitions of $R_*$ and $\bar\rho$ become tricky with the existence of the extended thin envelope bound to the star. We find that at the end of relaxation phase, the density at the original surface $R=1.03\,\Rsun$ is $\approx$$10^{-3}$\,g\,cm$^{-3}$. We therefore define $R_*$ as the radius at which the average density is $10^{-3}$\,g\,cm$^{-3}$ and evaluate $\bar\rho$ correspondingly. Additionally, tidal oscillations break the spherical symmetry of the star, and \citet{Ryu_2020c} found that the surviving star can be substantially oblate. To account for the fluctuation at the surface, we evaluate both the mean and standard deviation of $\rho$ at a sequence of distances from the center, so we can define a confident interval of $R_*$ within which the density profile is in 1$\sigma$ consistence with $10^{-3}$\,g\,cm$^{-3}$. 

In Figure~\ref{fig:deltaM} we show the mass loss $\Delta M$ during each passage for all three models. The uncertainty of $\Delta M$ is given by the marginally bound mass (typically $\approx$$10^{-4}$--$10^{-3}$\,$\Msun$), which we define as bound material ($E_*<0$) outside $R_*$ ($\rho<10^{-3}\,\mathrm{g\,cm^{-3}}$). In all three models, as the star is continuously tidally heated, $\xJLS$ increases over time, and the star gets progressively more vulnerable to disruption, with $\Delta M$ going up roughly exponentially. {Stars on orbits of a larger $\binit$ would undergo a more rapid increase in $\Delta$M over multiple orbits. On average, $\Delta M$ increases by a factor of $\sim$$1.3$, $1.8$, and $2.6$ between consecutive pericenter passages for $\binit=0.5$, $0.6$, and $1.0$, respectively. Assuming $\Delta$M keeps growing at the same pace, the star on the $\binit=0.5$, $0.6$ orbits will lose most of its original mass in $\sim$eight and two more orbits, {respectively}.} In the right panel of Figure~\ref{fig:deltaM} we overplot targets in the Stellar Tidal Disruption Events with Abundances and Realistic Structures (STARS) library from \mesa\ main-sequence models of a wide range of stellar masses, ages, and orbital parameters \citep{Law-Smith2020}. Despite some chaos due to tidal modes excited in previous passages not being fully damped, our results still agree well with the trace of the STARS main-sequence stars. This confirms that $\rhocrhobar$ is an ideal summary quantity in evaluating $\Delta M/M_*$, regardless of the star's mass loss and/or tidal interaction history.

\begin{figure*}
    \centering
    \includegraphics[width=\linewidth]{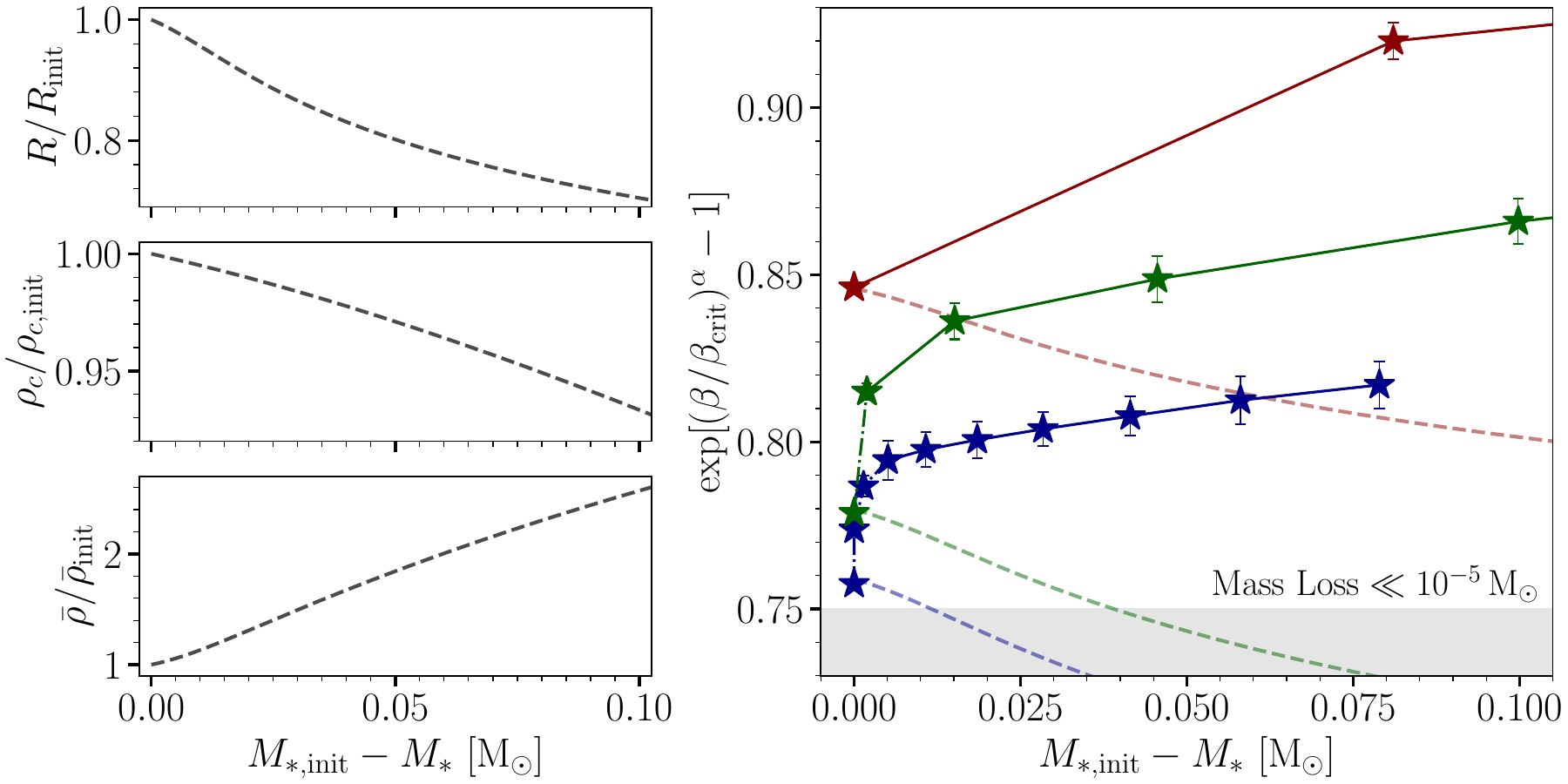}
    \caption{Models assuming adiabatic mass loss dramatically underestimate the vulnerability of the star to upcoming tidal disruptions. {\it Left:} the evolution of the stellar radius $R_*$, central density $\rho_c$, and average density $\bar\rho$ in a \mesa\ model, when we adiabatically remove 0.1\,$M_\odot$ from a sun-like star via a constant stellar wind. {\it Right:} $\xJLS$ as a function of the total mass loss. The results from our hydrodynamic simulations are plotted again as pentagrams, while the dashed lines are how we expect the star to evolve on orbits of corresponding $\beta_\mathrm{init}$ using the results from the adiabatic model. The dashed-dotted lines again indicate overestimation of the tidal energy injection due to numerical dissipation in our simulations. These adiabatic tracks should not go below $\xJLS\simeq0.75$, when significant mass loss will not happen (see Figure~\ref{fig:deltaM}).}
    \label{fig:mesa}
\end{figure*}

In Figure~\ref{fig:mesa} we show the evolution of $\xJLS$ as the star loses mass in our hydrodynamic simulations and in the \mesa\ model which only takes into account adiabatic mass loss. In the left panels we show the adiabatic evolution of $R_*$, $\rho_\mathrm{c}$, and $\bar\rho$ of the sun-like star. $R_*$ shrinks as the star is adiabatically stripped and $\rhocrhobar$ drops over time, consistent with analytical results \citep[e.g.,][]{Hjellming_1987, Dai_2013}. The net effect is a decreasing $\xJLS$, meaning that while the star is getting less concentrated, its tidal radius shrinks even more rapidly. When $\xJLS$ drops below $\approx$0.75, the star is effectively detached from the SMBH and the mass loss ceases. This trend is the opposite to what we find in hydrodynamic simulations, suggesting that tides play a critical role in injecting energy and reshaping the density structure of the star. 

In Section~\ref{sec:tidal_dis} we suggest that the number of orbits before a star reaches $R_{\Delta L}$ is chaotic and dramatically underestimated in our simulations (indicated as dashed lines in Figure~\ref{fig:deltaM}). For our $\binit=0.5$ and 0.6 models, $R_*$ exceeds $R_{\Delta L}$ in three and one orbit(s). The number of actual orbits can be roughly estimated under an energy consideration. The mass loss when the star reaches $R_{\Delta L}$ is still minimal ($\lesssim$1\%). Neglecting order unity coefficients, the increase in the total energy is
\begin{equation}
    \Delta E_* \simeq \frac{G\mathrm{M^2_\odot}}{\Rsun}\left(1 - \frac{\Rsun}{R_{\Delta L}}\right).
\end{equation}
Assuming the tidal oscillations are damped efficiently, the number of orbits it requires to inject this amount of tidal energy is
\begin{equation}\label{eq:n_orb}
    N_\mathrm{orb}(R_{\Delta L}) \simeq \frac{\Delta E_*}{\delta E_*} \approx \beta^{-6}T^{-1}(\beta)\left(1 - \frac{\Rsun}{R_{\Delta L}}\right).
\end{equation}
This corresponds to $\approx$$10^3$ and $\approx$$10^2$ orbits for $\binit=0.5$ and 0.6, respectively. 
As the star expands beyond $R_\mathrm{\Delta L}$, we are able to model the last few orbits when we expect most of the mass loss and the strongest flares. In Section~\ref{sec:flare} we will draw a sketch of the repeating flares as the stars step to the ends of their lives.

\section{Implications for Flare Evolution}\label{sec:flare}

\begin{figure*}
    \centering
    \includegraphics[width=\linewidth]{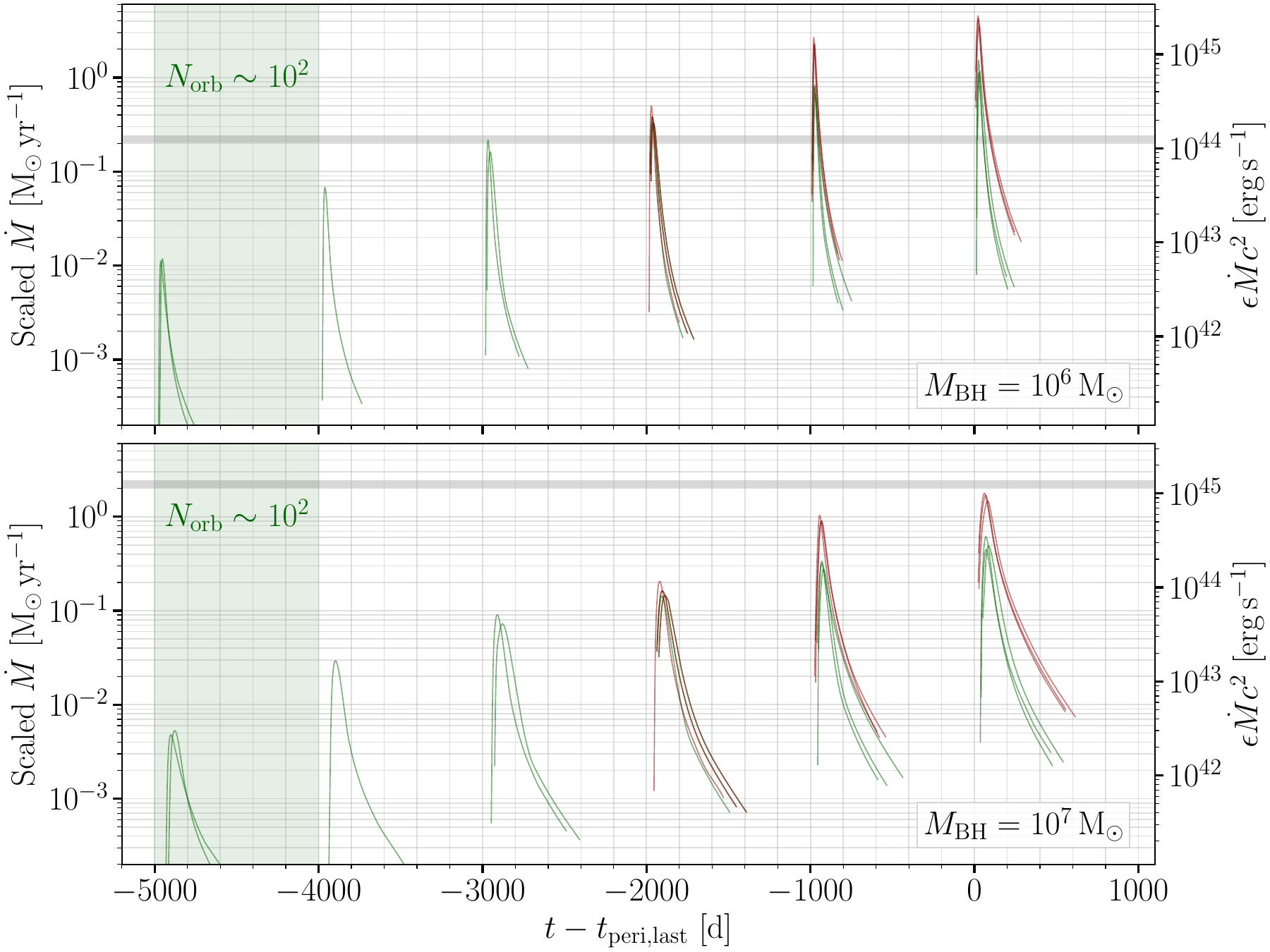}
    \caption{An illustration of mass fallback rates $\dot M(t)$ over flares for the $\beta_\mathrm{init}=0.6$ and 1.0 models, assuming two different $\mbh$. The time is measured with respect to the last pericenter passage {in our simulation, which is expected to produce the strongest flare}. For each passage, we adopt fallback curves from their $\le$three nearest neighbors in the $\log(\Delta M/M)$ - $\xJLS$ space from  the STARS library (the gray circles in Figure~\ref{fig:deltaM}). When adopting those mass fallback curves, we assume neighbors would show similar $\dot M/M_*$, then scale $\dot M$ using the mass of the star before each passage in the hydrodynamic simulations. We overlay the mock luminosity, assuming a radiation efficiency $\epsilon=10^{-2}$ on the right. The horizontal dashed lines mark the Eddington luminosities of SMBHs of corresponding masses. We note that the tidal heating in the first passage of the $\binit=0.6$ model is overestimated, and there could be $\sim$$10^2$ progressively stronger flares (the luminosity increases by a factor of $\sim$10).}
    \label{fig:fallback}
\end{figure*}

In our zoom-in simulations, debris tails, which can expand up to $\gtrsim$$10^2\,\Rsun$ from the surviving star before $\md M/\md E$ settles down, are not modeled. We thus cannot produce $\dot M$ directly from our simulations. Nevertheless, it has been shown that similar to $\Delta M/M_*$, the normalized mass fallback rate at peak $\dot M_\mathrm{peak}/M_*$ is also strongly correlated with $\xJLS$ for main-sequence stars \citep{Law-Smith2020}. We therefore expect the same relation should apply for strongly perturbed stars. As an illustration of last few flares as the star approaches the full disruption, in Figure~\ref{fig:fallback} we show the corresponding mass fallback curves from $\le$three nearest neighbors in the $\log(\Delta M/M_*)$ - $\xJLS$ space from  the STARS library for each of our passages. For neighbors, we allow deviations in $\log(\Delta M/M)$ and $\xJLS$ of no more than 0.15 and 0.02, respectively. We expect the $\dot M/M_*$ of these objects adjacent to the tidally reshaped star in the phase space to appear similar themselves and mimic the actual mass fallback curve. Simulations with low $\beta$ are rare in the STARS library, and the $\binit=0.5$ models do not have sufficiently nearby neighbors in phase space. We thus do not present them in the plot.

We consider two SMBH masses ($10^6$ and $10^7\,\Msun$) and an orbital period $P_\mathrm{orb}=1000$\,days, comparable with AT\,2020vdq which has one of the longest periods among repeating TDE candidates. While results in STARS library use simulations on a parabolic orbit around a $10^6\,\Msun$ SMBH, the $\md M/\md E$ distribution can be scaled to disruptions on highly eccentric \citep[$e\gtrsim0.9$;][]{Liu2023} orbits around a wide range of SMBHs \citep{Lodato_2009, Ramirez-Ruiz_2009, Haas_2012, Guillochon_2013}, as long as the encounter is nonrelativistic.\footnote{For a sun-like star with $\beta\lesssim1.0$, $\mbh$ cannot exceed $\approx$$10^7\,\Msun$ for $r_\mathrm{p}>10\,r_g$ to hold.} Following the approach illustrated in \citet{Liu2023}, we scale the $\md M/\md E$ in the STARS library and calculate $\dot M$. 

For each passage of both $\binit$, the mock mass fallback curves of their neighbors indeed show similar $\dot M_\mathrm{peak}$. This suggests that with similar $\xJLS$, a similar fraction of bulk mass falls back at a comparable timescale. The shapes of the curves, which depend sensitively on the fine structure of $\md M/\md E$, appear heterogeneous, so we cannot constrain the long-term evolution of, e.g., rise and decay timescales in repeating TDEs.

We expect an approximately exponential growth of $\dot M_\mathrm{peak}$ in the last few flares. If the radiation efficiency $\epsilon$ does not significantly change, the repeated disruption of a sun-like star on a $\binit=1$ orbit would brighten by a factor of $\approx$10 in its last $\approx$three flares, and a factor of $\approx$20 in the last $\approx$five flares for the same star on a $\binit=0.6$ orbit. On the other hand, early on in the evolution of a star on a low $\binit$ orbit, tides excited by the SMBH are much weaker, so $\Delta M$ and $\dot M$ should evolve much slower. If $\binit\lesssim0.6$, a star could produce over $10^2$ weak flares peaking below $\approx$$10^{-2}\,\Msun\,\mathrm{yr^{-1}}$, losing $\lesssim$$10^{-3}\,\Msun$ in each orbit. By adopting a typical $\epsilon=10^{-2}$ \citep{Dai_2013, 2018ApJ...859L..20D, Jiang_2016, Mockler_energy_2021}, the corresponding peak luminosity of these flares are well below the Eddington luminosity $L_\mathrm{Edd}$ ($\approx$$10^{-1}$ or $10^{-2}\,L_\mathrm{Edd}$ for $\mbh=10^6$ or $10^7\,\Msun$; Figure~\ref{fig:fallback}). A significantly shorter $P_\mathrm{orb}< 1\,$yr would squeeze the fallback timescale and boost the $\dot M_\mathrm{peak}$, producing sharper, brighter flares \citep{Liu2023}. {We note that, recently, \citet{Bandopadhyay_2024} simulated the flares produced by main-sequence and evolved stars over multiple partial disruptions using smoothed particle hydrodynamics. In the case of a 1\,$\Msun$ main-sequence star that we both have explored, our results are qualitatively consistent with each other.}

Since we are biased toward this last flare in flux-limited surveys (see Section~\ref{sec:pop} for more discussion), with the luminosity of one flare only, we cannot constrain either the orbital parameter $\beta$ or the stellar structure -- the star can be an unperturbed one on its first (and last) encounter with $\binit\gtrsim\beta_\mathrm{crit}$ or a strongly perturbed one after a couple of, dozens of, or even hundreds of weaker encounters. An underlying population of repeating TDEs could thus add to the difficulty of probing progenitor properties using light curves of a single TDE. Future large-scale simulations are needed to characterize the shapes of mass fallback curves in repeating TDEs.

\section{Discussion}\label{sec:discussion}
\subsection{Diverse Evolution Tracks from the Zoo of Disrupted Stars}\label{sec:star_zoo}

In this work, we focus on the structure of a sun-like star over partial tidal encounters. Similar to sun-like stars, upper-main-sequence stars $\gtrsim$1\,$\Msun$ have a deep radiative zone with a positive entropy gradient and would contract in response to adiabatic mass loss \citep[e.g.,][]{Soberman_1997}. Low-mass stars, which should dominate the stellar population by number in the initial mass function, are fully convective and react to mass loss by adiabatic expansion. Consequently, they quickly become much more vulnerable in the consecutive tidal encounter. The amplitude of tidal perturbation also depends sensitively on the stellar structure, and fully convective stars gain much more tidal energy in a tidal encounter at the same distance \citep{Lee_1986, Ivanov_2001}. In combination, low-mass stars have a much smaller $\beta_\mathrm{crit}$ for full disruptions than upper-main-sequence stars \citep{Guillochon_2013, Mainetti_2017, Law-Smith2020, Ryu_2020b} and, as repeaters, should survive fewer orbits with a more dramatic increase of $\Delta M$.

As stars evolve, a condense core is developed, which helps the star to retain more mass in tidal encounters \citep{Liu_2013}. In response to mass loss, the tenuous envelope of an evolved star would first inflate as a $\gamma=5/3$ polytrope until a substantial amount of mass is lost, such that the gravity of the core takes over. Consequently, $\Delta M$ would start to decrease following the contraction of envelope (with a nearly constant mass loss rate around the turning point), and the star could survive substantially more tidal encounters \citep{MacLeod_spoon_2013, Liu2023, Bandopadhyay_2024}, potentially leaving a hydrogen-depleted core behind \citep{Bogdanovic_2014}.

A diverse stellar diet for SMBHs could lead to miscellaneous flare properties. We have shown the episodic disruption of a sun-like star would produce a series of exponentially brighter flares -- the subsequent flare is $\approx$two to four times brighter than the previous one -- following many weaker flares that only mildly increase in luminosity. For a low-mass star, we expect more dramatic leaps over flares. A giant increase in luminosity over flares is observed in AT\,2020vdq \citep{Somalwar_2023}, in which the second flare is a factor $\gtrsim$10 more luminous,\footnote{The luminosity of the first flare is estimated with optical fluxes only \citep{Somalwar_2023} and is largely uncertain.} consistent with the picture of a repeating disruption of a main-sequence star. For evolved stars with a massive core, we expect slower evolution over flares or even a decreasing luminosity. ASASSN-14ko, the only repeating TDE with more than a handful of flares observed, shows no significant long-term trend in over 20 outbursts \citep{Payne_2021, Payne_2022} and is more consistent with an evolved star.\footnote{While a sun-like star on a $\binit\lesssim0.6$ orbit could also produce a series of flares of nearly constant luminosity before it expands substantially, the tiny $\Delta M/M_*\lesssim10^{-3}$ is not sufficient to power the flares observed \citep[$\Delta M\simeq0.04$--0.08\,$\Msun$;][]{Liu2023}.} Future hydrodynamic simulations will quantify the different evolution tracks for a variety of stellar masses and ages.

\subsection{The Edge of the Loss Cone}\label{sec:loss_cone}

An underlying assumption in our analysis is that the stellar orbit does not significantly change over multiple orbits, which is only possible in the empty-loss-cone regime, when the relaxation timescale $t_\mathrm{relax}$ of the orbital angular momentum is much greater than $P_\mathrm{orb}$ \citep{Stone_2020}. In this limit, stars are nearly adiabatically scattered to an orbit of the smallest possible $\binit$ leading to a full disruption following a series of weak disruptions \citep{Bortolas_2023}, where $\binit$ can be significantly less than 1. We expect most of the repeating TDEs occur at this boundary $\binit$, thus it is critical in evaluating TDE rates. Given the diverse stellar diet for SMBHs, it also depends on stellar types.

Within $t_\mathrm{relax}$, two other effects dominate the orbital variation: tidal excitation converts the orbital energy of the star to its internal energy, pulling it to a tighter orbit, whereas asymmetric mass loss kicks the star away \citep{Gafton_2015, Kremer_2022}. 
\cite{Broggi_2024} studied these competitive effects jointly while neglecting the change in the stellar structure, and showed that mass loss kicks always dominate in a $\gamma=4/3$ polytrope (upper main sequence), pushing the star to a less bound orbit. As the angular momentum is nearly fixed, the star migrates toward a lower $r_\mathrm{p}$. Consequently, in the empty-loss-cone regime, for upper-main-sequence stars, the edge of the loss cone is roughly set by the smallest $\beta$ with $\Delta M/M_*\simeq q$, where $q\simeq P_\mathrm{orb}/t_\mathrm{relax}$ is the loss cone filling factor,\footnote{The $\Delta M/M_*$ we derive for $\beta\lesssim0.6$ is much lower than what Equation~(5) in \citet{Broggi_2024} predicts, probably due to an extrapolation in $\beta$ since \citet{Ryu_2020c} did not perform simulations below $\beta=1$ for sun-like stars.} 
which guarantees a full disruption within $t_\mathrm{relax}$. For $\gamma=5/3$ polytropes (low-mass stars), the orbital variation is dominated by tidal excitation instead, so they would migrate to orbits with larger $r_\mathrm{p}$. As a result, the minimal $\binit$ typically needs a $\Delta M/M_*>q$. We note that the evolution of the structure and orbit of a star driven by both mass loss and tidal excitation has never been studied coherently.
More realistic simulations of the star's response to weak tidal encounters are of particular importance.

\subsection{Constraining the Population of Repeating TDEs}\label{sec:pop}
The volumetric event rate and population-level properties (e.g., $P_\mathrm{orb}$ and $\binit$) for repeating partial TDEs are highly uncertain. Nevertheless, partial TDEs on long-period orbits could dominate the entire TDE population with two-body relaxation only \citep{Bortolas_2023} or aided by eccentric Kozai-Lidov effects in SMBH binaries \citep{2023ApJ...959...18M,Melchor_2024}. Other mechanisms will be needed to produce the population of observed repeaters with ultrashort $P_\mathrm{orb}\lesssim1\,\mathrm{yr}$ \citep[e.g., the Hills mechanism;][]{Cufari_2022, Lu_2023}. 

For long-period repeating TDEs, where $P_\mathrm{orb}$ is beyond the lifetime of a survey (or us astronomers), they are likely not distinguishable from nonrepeaters observationally. For these repeaters, our surveys can be heavily biased toward the last few flares. We have shown that a sun-like star could spend numerous orbits before any significant radius expansion or mass loss. In the case of $\binit=0.6$ we expect the star to spend most of its $N_\mathrm{orb}\simeq10^2$ orbits losing a minimal amount of mass, before releasing most of its energy in the last {$\approx$five} flares that are $\gtrsim$$N_\mathrm{orb}$ times more energetic. In a flux-limited survey, the volume of the Universe we could probe depends on the peak luminosity of a transient, $\mathcal{V}\propto L_\mathrm{peak}^{3/2}$. While the early-on weaker flares dominate in volumetric event rate by a factor of $\approx$$N_\mathrm{orb}$, the terminal flare dominates in the cumulative observed probability by a factor of $\approx$$N_\mathrm{orb}^{1/2}$. In evaluating TDE rates to reconcile with observations, the population of repeating TDEs will need to be carefully considered.

For short-period repeaters, as the objects discovered so far, a systematic all-sky survey can help us constrain their population properties. \citet{Somalwar_2023} performed a Monte Carlo test of repeating partial TDEs with $P_\mathrm{orb}$ spanning 0.3--2.7\,yr (between that of ASASSN-14ko and AT\,2020vdq) by searching for rebrightening in an optical TDE sample \citep{Yao_2023} from the Zwicky Transient Facility \citep[ZTF;][]{Bellm_ZTF_2019a}. With AT\,2020vdq being the only apparent repeater, an upper limit of the event rate for repeaters is $<$30\% the rate of the entire population, which will be less stringent if allowing for higher $P_\mathrm{orb}$. We note that an implicit hypothesis by looking for rebrightening is that the observed flare would not be the last one, which, given the observational bias we have addressed, leads to underestimation of rates. Searching for both faint preflares and rebrightening will alleviate the systematics, which requires a long, deep survey (readers are referred to the Appendix for a detailed study). The 10\,yr Rubin Observatory Legacy Survey of Space and Time (LSST), with extraordinary sensitivity, will provide a unique window of finding repeating partial TDEs systematically. By combining the legacy of Rubin with optical surveys of smaller aperture telescopes (e.g., BlackGEM and LS4), we will be able to explore both the luminous and the faint ends of repeating TDEs to obtain a better understanding the population underground. 

{
\subsection{Caveats}
At the end of this section, we note again that our implications for flare evolution are heuristic. Here, we summarize the major assumptions we have made and justify the potential uncertainty in our interpretation.

Above all, the spread in the binding energy within the tidal debris is approximated using the neighboring models in the $\log(\Delta M/M)$ - $\xJLS$ space from the STARS library, for which we are not able to predict the exact rise or decay timescale of each flare or any long-term trends. Nevertheless, in Figure~\ref{fig:fallback} we have demonstrated that the overall rising trend of the peak fallback rate is not smeared.

The mass fallback rate is then estimated assuming the tidal debris moves in the gravitational field of the SMBH, and the surviving star is neglected. It has been demonstrated in hydrodynamic simulations of partial TDEs, though, that within $10^2\,\tdyn$ ($\lesssim$1\,day) after the pericenter passage of a main-sequence star, the distribution of the orbital binding energy within the tidal debris is settled  \citep{Law-Smith2019}. Both the gravitational and hydrodynamic impacts of the surviving star have mostly ceased by then. For repeating TDEs on highly eccentric orbits with $P_\mathrm{orb}\gg 1$\,day, this approximation is solid. For less eccentric orbits and/or less extreme mass ratios (e.g., stars disrupted by stellar-mass or intermediate-mass BHs on mildly eccentric orbits), the surviving star would play a more important role in the fallback of the tidal debris.
    
To convert the mass fallback rate into the accretion rate and thus the luminosity, we assume prompt disk formation and ignore the interaction between the surviving star and the accretion flow. It is known that the tidal debris cannot circularize rapidly in mildly eccentric TDEs: (i) the spread in the orbital energy in the tidal tail is significantly reduced in eccentric TDEs, so the compression and shearing at the pericenter will be less efficient in dissipating the orbital energy; and (ii) the self-interaction in the debris stream induced by the general relativity apsidal precession will happen closer to the SMBH with a lower relative velocity. For $e=0.8$, the tidal debris would not circularize for more than $10\,P_\mathrm{orb}$ of the star \citep{Bonnerot_2016, Hayasaki_2016}, which may be further delayed by the Lense-Thirring frame dragging of a rapidly spinning SMBH. In this case the surviving star will surely interact with the material stripped over multiple previous tidal encounters near the pericenter, which may both alter the star's orbit and reshape the accretion flow. 

The circularization is much more efficient in highly eccentric (which are probably more realistic) repeating TDEs (for a sun-like star, an orbit with $P=100$\,days and $\beta=1$ has an $e\simeq0.99$), and should be comparable to that of canonical parabolic TDEs \citep{Bonnerot_2016}. In addition, the viscous dissipation at the typical circularization radius of $2\,r_\mathrm{p}$ is rather efficient. For a sun-like star getting inflated in multiple pericenter passages, we still expect $t_\mathrm{vis}\lesssim10^{-2}P_\mathrm{orb}$ if $P_\mathrm{orb}>100$\,days \citep{Liu2023}. As long as the disk formation is prompt, the mass would be rapidly accreted, when the interaction between the star and the accretion flow is less important. This is likely the case in the first 17 observed flares of ASASSN-14ko \citep{Payne_2021}, all of which showed a rapid rise consistent with a negligible viscous delay. Two recent flares of ASASSN-14ko, however, evolved much slower, with a bump on the rise \citep{Huang_2023}, which might indicate the presence of a slowly accumulating accretion disk that only recently started to interact with the infalling tidal debris.

In summary, the long-term behavior of the material stripped in multiple tidal encounters remains a major unknown in our study. Future hydrodynamic simulations of the stripped material in the joint, time-dependent gravitational field of the SMBH and the surviving star will be critical to resolving these unknowns.
}

\section{Conclusions}\label{sec:conclusion}
We have presented  hydrodynamic simulations of a sun-like star over multiple tidal encounters with an SMBH and provided observational implications. The star will be significantly reshaped at each pericenter passage, due to both mass loss and tidal heating, becoming more vulnerable in subsequent encounters. Consequently, the star is doomed to a full disruption after producing a (potentially) large number of weak flares, followed by a couple of luminous flares with an exponential increase in brightness. Stars with a different initial density structure (e.g., evolved stars with a massive core) may follow distinct evolution tracks, which needs to be quantified in future numerical studies.

We confirm the correlation between $\Delta M/M_*$ and $\xJLS$ (a function of $\rhocrhobar$ and $\beta$) for main-sequence stars also applies for strongly perturbed stars surviving multiple tidal encounters. This degeneracy adds to the difficulty of distinguishing full TDEs on the star's first tidal encounter from the last dance after a series of weaker encounters, possibly limiting our ability to infer the stellar properties using the light-curve information of a single flare.

Repeating flares from weaker encounters can be systematically missed due to the limited instrumental sensitivity and survey span (below the recurrence time). Rubin provides the chance to explore the faint end of the luminosity function of TDEs and the event rates of repeaters, opening up a unique window for probing the dynamics in the vicinity of SMBHs.

We thank the anonymous referee for constructive comments and suggestions that helped to improve the manuscript. We thank Katie~Auchettl, Brenna~Mockler, Ryan~Foley, Adam~Miller, and Meng~Sun for useful discussions. We are particularly grateful to James~Guillochon and Jamie~Law-Smith for  previous collaborations and guidance with the numerical setup. We thank the Heising-Simons Foundation, NSF (AST-2150255 and AST-2307710), Swift (80NSSC21K1409, 80NSSC19K1391) and Chandra (22-0142) for support. 
The ZTF forced-photometry service was funded under the Heising-Simons Foundation grant \#12540303 (PI: Graham).

\software{\flash\ \citep{Fryxell_FLASH_2000}, \mesa\ \citep{Paxton2011, Paxton2013, Paxton2015, Paxton2018, Paxton2019, Jermyn2023}, \texttt{yt} \citep{yt_2011}}

\bibliography{Software,TDE,Star}
\bibliographystyle{aasjournal}

\appendix
\section{Constraining Preflares of TDEs Using Archival Data}\label{sec:obs}
Here, we show examples of two ZTF TDEs to illustrate the difficulties in constraining preflares with archival ZTF forced-photometry data \citep{ZTF_FP_2019}. ZTF20acqoiyt is a bright TDE in ZTF's bright transient survey \citep{BTS_2020}, whereas ZTF21abxngcz is one of the most distant ($z=0.2860$) and luminous ($L\simeq10^{45}\,\mathrm{erg\,s^{-1}}$) objects in the sample of \citet{Yao_2023}. Given that ZTF21abxngcz is over five times more distant than ZTF20acqoiyt, it is 1.2\,mag fainter in $g$ at the peak, despite its high luminosity. While the historical nondetections at the position of ZTF20acqoiyt are sufficiently deep to rule out a preflare that is 30\% (typical from our simulations) or 10\% (AT\,2020vdq-like) as bright as the primary flare in the past 2.4\,yr, for ZTF21abxngcz, an AT\,2020vdq-like preflare could only be marginally ruled out. Seasonal observing gaps and variations in the detection limit (nonideal airmass, weather, or the Moon) add additional difficulties in ruling out preflare existence. The sensitivity of Rubin/LSST is of particular importance in searching for these faint preflares.
\begin{figure}[h!]
    \centering
    \includegraphics[width=0.5\linewidth]{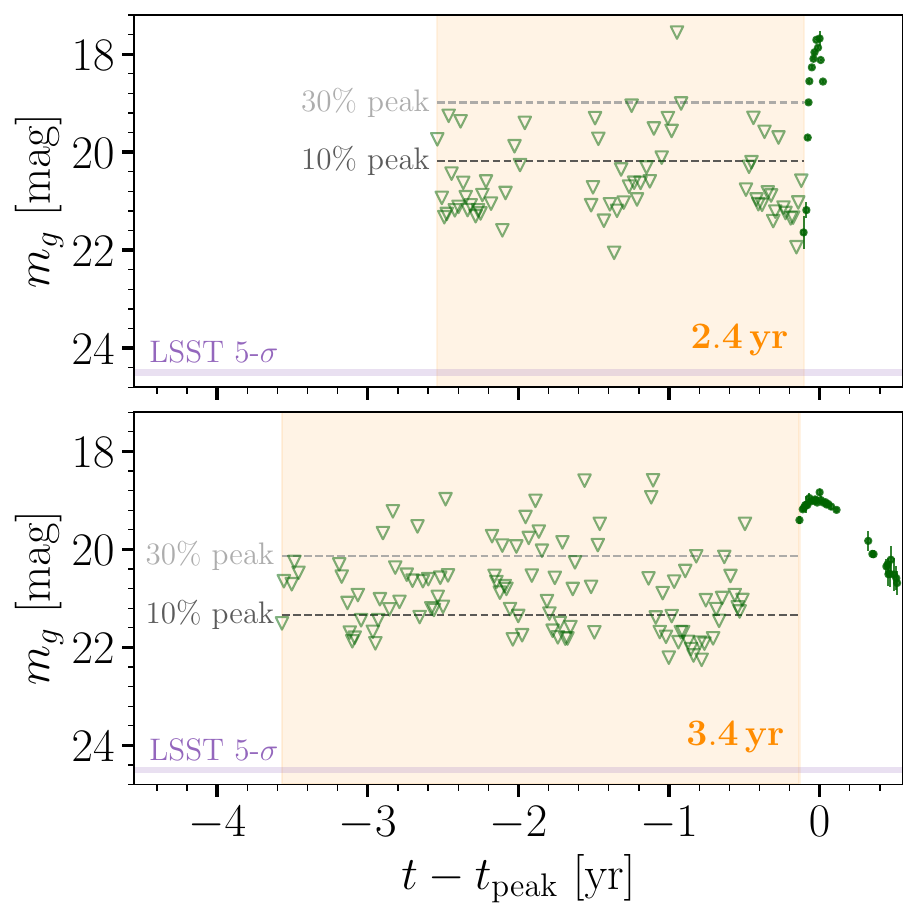}
    \caption{Forced-photometry light curves of two ZTF TDEs (ZTF20acqoiyt and ZTF21abxngcz) in $g$, where the green triangles indicate 5$\sigma$ upper limits of 3$\sigma$ non-detections, which have been stacked into 5\,day bins to improve the signal to noise ratio. The orange shaded regions mark the span between the first ZTF observation and the first 3$\sigma$ detection, setting the longest $P_\mathrm{orb}$ we could possibly constrain. The horizontal dashed lines indicate where a hypothetical preflare 30\% (typical in our simulations) or 10\% (AT\,2020vdq-like) as bright as the primary flare would peak.}
    \label{fig:ZTF_TDE}
\end{figure}

\end{CJK*}
\end{document}